**Development of a Validation and Inspection Tool for Armband-based Lifelog Data (VITAL) to Facilitate the Clinical Use of Wearable Data: A Prototype and Usability Evaluation**

Eunyoung Im, RN, MPH [1, 2]; Sunghoon Kang, MS [3]; Hyeoneui Kim, RN, MPH, Ph.D., FAAN [1, 2, 4]*

[1]College of Nursing, Seoul National University, Seoul, Republic of Korea

[2]Center for World-leading Human-care Nurse Leaders for the Future by Brain Korea 21 (BK 21) four project, College of Nursing, Seoul National University, Seoul, Republic of Korea

[3]College of Natural Sciences, Seoul National University, Seoul, Republic of Korea

[4]The Research Institute of Nursing Science, Seoul National University, Seoul, Republic of Korea

*Corresponding author:

Hyeoneui Kim

College of Nursing, Seoul National University

103 Daehak-ro, Jongno-gu, Seoul 03080, South Korea

Tel) +82-2-740-8483

Email: ifilgood@snu.ac.kr

# Development of a Validation and Inspection Tool for Armband-based Lifelog Data (VITAL) to Facilitate the Clinical Use of Wearable Data: A Prototype and Usability Evaluation

## Abstract

**Background:** The rise of mobile technology and health apps has increased the use of person-generated health data (PGHD). PGHD holds significant potential for clinical decision-making but remains challenging to manage.

**Objective:** This study aimed to enhance the clinical utilization of wearable health data by developing the Validation and Inspection Tool for Armband-Based Lifelog Data (VITAL), a pipeline for data integration, visualization, and quality management, and evaluating its usability.

**Methods:** The study followed a structured process of requirement gathering, tool implementation, and usability evaluation. Requirements were identified through input from four clinicians. Wearable health data from Samsung, Apple, Fitbit, and Xiaomi devices were integrated into a standardized dataframe at 10-minute intervals, focusing on biometrics, activity, and sleep. Features of VITAL support data integration, visualization, and quality management. Usability evaluation involved seven clinicians performing tasks, completing the Unified Theory of Acceptance and Use of Technology (UTAUT) survey, and participating in interviews to identify usability issues.

**Results:** VITAL successfully integrated wearable data, thus enabling all participants to complete tasks with minimal errors without prior participant training. UTAUT survey results were positive, with average scores of 4.2 for performance expectancy, 3.96 for effort expectancy, and 4.14 for intention to use, indicating high user satisfaction and intent to adopt the tool.

**Conclusions:** By enhancing wearable data integration, visualization, and quality management, the VITAL prototype shows significant potential for clinical application. Positive feedback highlights its promise, while emphasizing the need for further studies to confirm its real-world effectiveness.

# Introduction

The advancement of mobile technology is revolutionizing the healthcare sector, with mHealth technologies playing a crucial role [1]. Due to the burgeoning awareness of health and wellness among the general population, the market size of mobile health applications and wearable devices as tools for healthcare is increasing [2,3]. These tools can collect various health-related data such as heart rate, oxygen saturation, step count, calories burned, exercise, and sleep patterns. The data created, recorded, or gathered by patients to self-manage their health are known as person-generated health data (PGHD) [4], and health data from wearable devices form a major portion of it. Health data collected from wearable devices are increasingly valued for their potential to monitor patient health conditions and provide tailored feedback [5]. Moreover, the use of wearable health data can enhance therapeutic communication between clinicians and patients [6,7] and improve the treatment outcomes for patients with chronic conditions [8,9]. Clinicians acknowledge the clinical utility of wearable health data and encourage patients to collect the data to gain a better understanding of their health status outside of the hospital setting [10-12]. In addition, patients believe that the wearable health data they collect should be utilized by healthcare professionals during clinical encounters [13,14].

However, despite the anticipated advantages it brings to patient care, its clinical use remains challenging. For healthcare professionals, reviewing wearable health data is a significant burden in practice due to the overwhelming volume of data, the technological complexities associated with handling and integrating data, and concerns related to data quality [15-18]. Assuredly, clinical use of wearable health data was identified as a factor that exacerbates healthcare provider burnout in previous studies [9,18,19]. As a result, some clinicians hesitate to incorporate wearable data into the clinical process [18]. To address this issue, summarizing and visualizing data can make it more intelligible and provide quick insights, which can help alleviate the burden [18,20,21]. Therefore, to improve the implementation of wearable health data in clinical settings, it is essential to develop user-friendly systems that ensure data quality and enable meaningful data exploration [19].

Due to the absence of standardization, the lack of interoperability across devices is another challenge in using wearable health data in clinical settings [22,23]. This issue is a significant hindrance to the adoption of wearable data as a systematic component of clinical practice [24]. To be effective for use in a clinical setting, wearable health data must be processed and integrated into a standard format to enable data exploration and analysis. However, the lack of regulatory frameworks and variations in device types, data formats, and sampling rates complicates this process, impeding effective management and integration [22,25]. The successful incorporation of wearable health data in clinical settings necessitates the critical prerequisite of integrating diverse data elements into a consistent format [26]. The establishment of a data integration process that addresses these inconsistencies is essential for supporting the effective management and utilization of integrated wearable data in clinical contexts [27-29]. However, few studies have provided practical methods or concrete examples for processing heterogeneous wearable health data.

The issue of data quality is the most significant concern regarding the utilization of wearable health data. A significant number of clinicians were wary of the accuracy and reliability of wearable data and expressed reluctance to use it in patient care [10,20,30,31]. Factors affecting the quality of wearable health data can emerge at any stage, from collection to utilization, necessitating processes to ensure the data's integrity such as data processing, cleansing, and management [32]. During the data collection stage, efforts to address issues can be challenging due to manufacturer-specific algorithms or variability among individual device users, but ensuring high-quality data during utilization remains crucial [22]. Studies utilizing wearable health data have attempted to address data quality issues, primarily through daily step count or wear-time-based data filtering [33-36]. However, inconsistencies in quality control methods applied to wearable health data may affect the reliability and reproducibility of study results [37].

This study developed a prototype wearable health data processing tool, the Validation and Inspection Tool for Armband-based Lifelog data (VITAL), and evaluated its usability to address the issues described above. Additionally, we explored the suitability of utilizing wearable health data in clinical settings through the VITAL prototype and gained preliminary insights into the factors, such as visualization and quality management, that support its implementation.

# Methods

## User Requirement Analysis

The development stage of VITAL began with requirement gathering, during which the authors identified the key components and necessary functionalities of the VITAL prototype. Four clinicians (one physician and three nurses) were recruited for the study. In July 2023, one-hour unstructured interviews were conducted to gather clinical insights. To facilitate their understanding of VITAL, the participating clinicians were presented with hypothetical scenarios of using wearable health data in clinical cases prior to the interviews. Preferences on wearable health data presentation and the essential key features for enhancing VITAL's usability and efficacy were identified. In addition, the concerns and anticipated benefits regarding the use of wearable health data in clinical practice were ascertained. The key insights obtained through the user requirement analysis are summarized in Table 1.

**Table 1. Clinicians' needs and preferences for VITAL prototype.**

| Category | Needs and preferences |
|---|---|
| Preferred Functionalities for Effective Data Visualization | • Visualization of data trends<br>• Selective filtering to view trends for specific data items when multiple items overlap<br>• Flexible options for valid data filtering<br>• Data interpretation and summarization<br>• Inter-patient comparison |
| Usability Features for System Convenience | • Comprehensive information on a single screen<br>• Intuitive user interface (easy and simple to operate)<br>• Convenient layout<br>• Personalized menu settings<br>• Integration with electronic medical records (EMR) |
| Concerns Regarding | • Data reliability |

| Category | Needs and preferences |
|---|---|
| Wearable Health Data Utilization in Clinical Practice | • Patient compliance<br>• Applicability in acute patient care |
| Expected Benefits Regarding Wearable Health Data Utilization in Clinical Practice | • Continuous monitoring (between outpatient visits or during ward admissions)<br>• Useful for health promotion (primary care, community)<br>• Utility in patient education<br>• Objectivity of data and potential for future clinical data utilization |

## Development of VITAL prototype

### *Data Source*

The VITAL prototype was primarily developed for data from armband devices widely used in Korea, including Samsung® (Watch 4), Apple® (Watch 7), Fitbit® (Charge 5), and Xiaomi® (Mi Band 7) models. To understand data types, sampling rates, and data formats, wearable data from each manufacturer were collected by the authors for 2-114 days. The data were extracted using the associated companion mobile apps. This study focused on clinically relevant data items, as presented in Table 2. A systematic review of the collected data revealed differences in the formats extracted by each manufacturer, including variations in timestamp, data types, and measurement units (Tables S1 and S2 in Multimedia Appendix 1). Furthermore, the sampling rates and durations varied for each data item across manufacturers. Therefore, the need for a standardized integration approach was identified as essential to ensure effective data processing.

**Table 2. Data items processed in VITAL prototype.**

| Data items | Manufacturers (companion apps) | | | |
|---|---|---|---|---|
| | **Galaxy (Samsung Health)** | **Apple (Health)** | **Fitbit (Fitbit)** | **Xiaomi (Zepp life)** |
| **Activity** | | | | |
| Start time | DATETIME (YYYY-MM-DD HH:MM:SS, ms) | DATETIME (YYYY-MM-DD HH:MM:SS + tz) | DATETIME (MM/DD/YY HH:MM:SS) | DATE / TIME |
| End time | | | Not collected | Not collected |
| Step count | INTEGER | INTEGER | INTEGER | INTEGER |
| Activity duration | INTEGER | End-start | Not collected | Not collected |
| **Exercise** | | | | (Consistent activity) |
| Start time | DATETIME (YYYY-MM-DD HH:MM:SS, ms) | DATETIME (YYYY-MM-DD HH:MM:SS + tz) | DATETIME (MM/DD/YY HH:MM:SS) | DATE / TIME |
| End time | | | Not collected | Not collected |
| Step count | INTEGER | Not collected | INTEGER | INTEGER |
| Exercise duration | INTEGER | End-start | INTEGER | End-start |
| **Heart rate** | | | | |
| Start time | DATETIME (YYYY-MM-DD HH:MM:SS, ms) | DATETIME (YYYY-MM-DD HH:MM:SS + tz) | DATETIME (MM/DD/YY HH:MM:SS) | DATE / TIME |
| End time | | | Not collected | Not collected |
| Heart rate | INTEGER | INTEGER | INTEGER | INTEGER |
| **Oxygen saturation** | | | | |
| Start time | DATETIME (YYYY-MM-DD HH:MM:SS, ms) | DATETIME (YYYY-MM-DD HH:MM:SS + tz) | DATETIME (YYYY-MM-DDT HH:MM:SS, ms) | Not extracted from Zepp Life |
| End time | | | Not collected | |
| Oxygen saturation | INTEGER | FLOAT | INTEGER | |
| **Sleep** | | | | |
| Start time | DATETIME (YYYY-MM-DD HH:MM:SS, ms) | DATETIME (YYYY-MM-DD HH:MM:SS + tz) | DATETIME (YYYY-MM-DDT HH:MM:SS, ms) | DATE / TIME |
| End time | | | Start + duration | Not collected |
| Sleep duration | INTEGER | End-start | INTEGER | INTEGER |
| **Sleep stage** | | | | |
| Start time | DATETIME (YYYY-MM-DD HH:MM:SS, ms) | DATETIME (YYYY-MM-DD HH:MM:SS + tz) | DATETIME (YYYY-MM-DDT HH:MM:SS, ms) | DATE / TIME |
| End time | | | Start + duration | Not collected |
| Sleep stage | INTEGER | String | String | String |
| Stage duration | INTEGER | End-start | INTEGER | INTEGER (Daily aggregated data) |

*Note.* ms: milliseconds; tz: timezone.

## Data Integration

As presented in Table 3, all labels, data formats, and measurement units for each data item were first standardized to ensure consistency and comparability across datasets. This standardization was essential for integrating data from multiple sources.

**Table 3. Standardization of data formats and measurement units.**

| Data items | Standardized data types (formats or measurement units) |
| --- | --- |
| Timestamp | DATETIME (YYYY-MM-DD HH:MM:SS) |
| Step count | INTEGER (count/min) |
| Activity/exercise duration | INTEGER (minutes) |
| Heart rate | INTEGER (beats per minute) |
| Oxygen saturation | INTEGER (percents) |
| Sleep stage | STRING (4 stages: deep, light, REM, awake) |
| Sleep duration | INTEGER (minutes) |

*Note*. REM: Rapid Eye Movement.

Next, the data were combined to facilitate integrative analysis by aligning measurement time intervals. To find the most suitable time interval for data integration, our study experimented with 1-minute, 5-minute, 10-minute, and 30-minute intervals. We then examined the differences in data density, heart rate distribution, and data processing time (Figure S1, Tables S1 and S2 in Multimedia Appendix 2). As a result, a 10-minute interval was deemed most effective for visualizing large volumes of data with minimal distortion.

The approach to data integration is illustrated by Figure 1. Figure 1 shows that data items were categorized into three types in terms of their collected timeframes in relation to the ten minutes windows. Activity, exercise, and sleep data were segmented and aggregated based on the pre-defined time windows. Activity or sleep data collected at shorter durations within 10-minute windows were summed (Figure 1-1), while data collected at a longer duration (over 10 minutes) were divided to fit the time window (Figure 1-2). Biometric data, such as heart rate and oxygen saturation, were averaged when multiple data points were included within a single time window (Figure 1-3). Data integration was performed for each data item, following the flow specific to each item (see Figure S1 in Multimedia Appendix 3). Figure 2 illustrates, after integrating data with a 10-minute interval, how the density of the data table with three data times changes. The integrated data were exported in CSV file format.

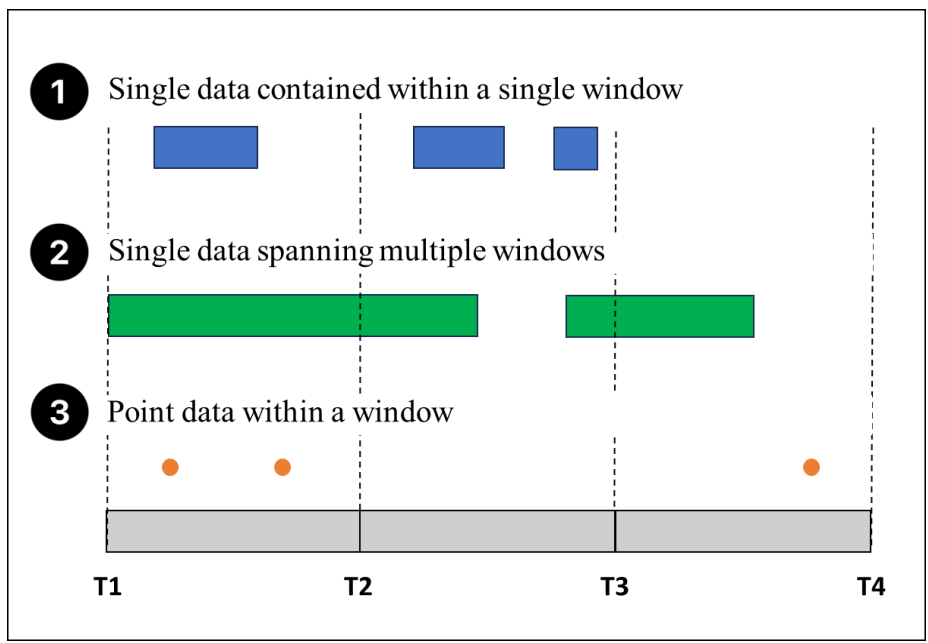

**Figure 1. Different data types capturing intervals and durations.**

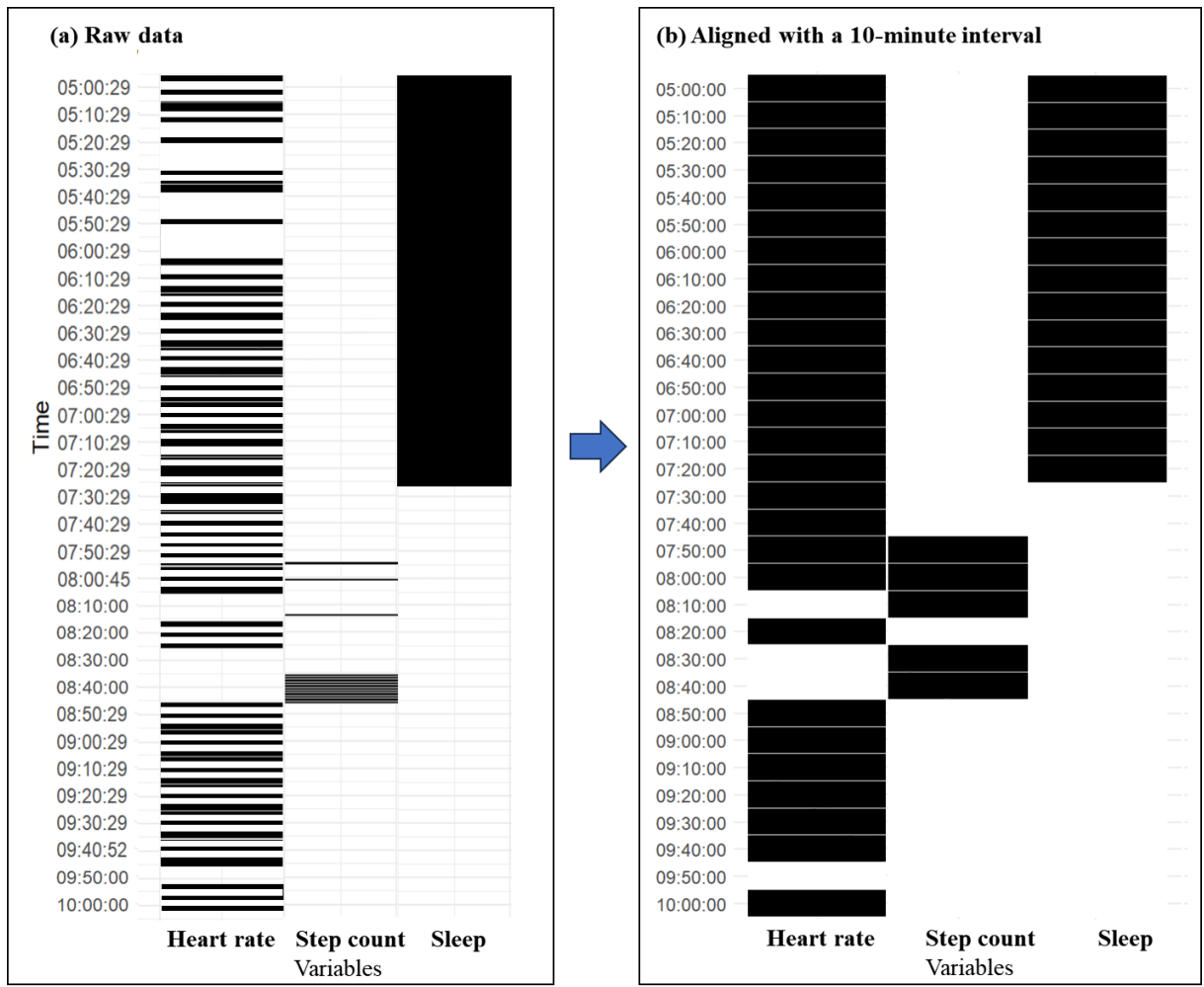

**Figure 2. Changes in data density after data integration.**

## Implementing the VITAL Prototype

The VITAL prototype was structured around two main functionalities: 1) a data integration module; and, 2) a visualization and quality management module. The data integration module of the VITAL prototype was implemented with HTML and Node.js, MongoDB, and Python. The visualization and quality management module was written in R (version 4.3.0) and R Shiny.

As described previously, the data integration module was designed to extract data items from the raw data files and integrate them into a standardized dataframe for effective visualization. Through graphs and textual summaries, the visualization and quality management module was developed to offer insights into data trends (Figure S2 in Multimedia Appendix 3). Furthermore, by applying filters, this module managed data quality to retain reliable data points. The filters incorporated daily total step counts and daily total device wear time, which were utilized by many studies to filter-in valid data [33,34,38-40]. The daily total step count was produced by combining the daily step records, and the device wear time was calculated by combining the time intervals during which any data item was recorded.

To measure data quality, we put three parameters into operation: completeness, recency, and plausibility. These parameters were identified by previous studies regarding quality dimensions for wearable data [22,23]. This study assessed completeness as the proportion of data present within each 10-minute timeframe throughout each day, with the average calculated from the start to the end of data collection. Presence was determined by the availability of any data from activity, biometrics, or sleep within a given timeframe. Recency was evaluated in two ways: (1) calculating the proportion of the collected data within a specified time period (e.g., recent one month); and, (2) determining the average age of the entire data. Plausibility involved assessing the relationship validity among the collected data items, such as examining whether step counts were recorded during sleep, checking for correlation between step counts and heart rate, and identifying overt outliers in heart rate data.

The beta version of VITAL prototype was presented to the clinicians who participated in the user requirements analysis to ensure that the clinical users' needs were sufficiently reflected in the tool. Overall, the clinical user requirements were deemed sufficiently reflected in the VITAL prototype.

## Usability Evaluation –VITAL prototype

The evaluation of VITAL prototype usability was conducted across three sessions (Figure 3). We recruited seven clinicians to participate. Usability issues were identified by observing participants performing tasks with the VITAL prototype in Session 1. Two tasks were carried out; usability metrics such as task success, error rate, and time per task were measured. The first task involved completing the uploading and data integration process. In the second task, participants were asked to freely explore the visualization module, apply quality management as described in the instructions, take a quiz, and subsequently provide feedback for a hypothetical patient. Session 2 involved administering the Unified Theory of Acceptance and Use of Technology (UTAUT) survey to assess participants' perceptions of performance expectancy (usefulness), effort expectancy, and their intention to use the VITAL prototype [41]. Session 3 consisted of individual interviews, during which participants discussed usability issues of the VITAL prototype, additional data quality metrics and criteria for practical use, and the clinical support needed for the effective use of wearable health data. All sessions were conducted without prior participant training.

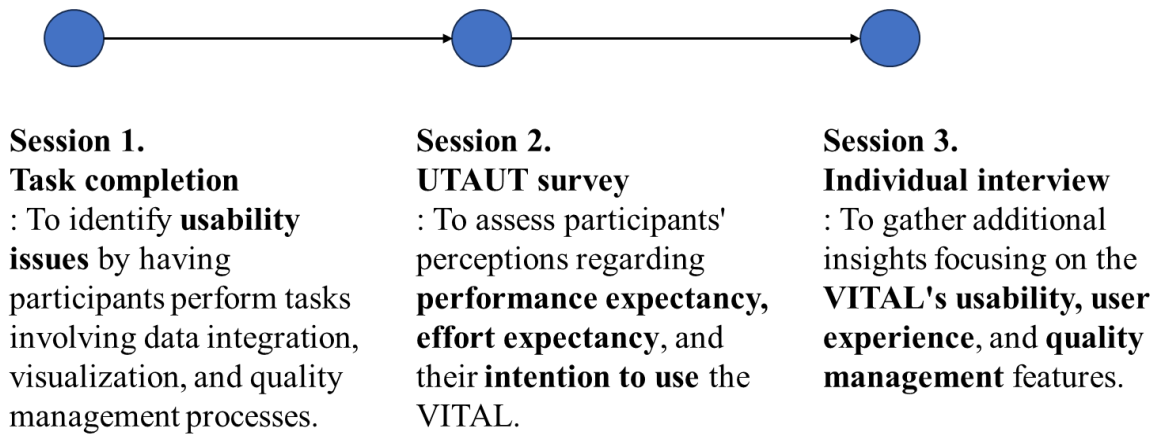

**Session 1.**
**Task completion**
: To identify **usability issues** by having participants perform tasks involving data integration, visualization, and quality management processes.

**Session 2.**
**UTAUT survey**
: To assess participants' perceptions regarding **performance expectancy, effort expectancy**, and their **intention to use** the VITAL.

**Session 3.**
**Individual interview**
: To gather additional insights focusing on the **VITAL's usability, user experience**, and **quality management** features.

**Figure 3. Usability evaluation process of VITAL prototype.**

## Ethical Considerations

This study was approved by the Institutional Review Board of the study site (requirement analysis: IRB No. 2308/001-006, tool evaluation: IRB No. 2312/001-015).

## Results

### VITAL Prototype Core Function

The overall VITAL pipeline is shown in Figure 4. Back-end processes manage data extraction and integration subsequent to user login and data upload to VITAL (Figure S3 in Multimedia Appendix 3). Data from each manufacturer and data item are integrated into a standardized data format. After integration, the data visualization and quality management module can be accessed by users for data exploration. Users can explore trends in patient wearable health data integrated at 10-minute intervals or aggregated daily (Figure 5). Furthermore, data quality can be managed based on reasonable criteria selected through the sidebar (red boxes in Figure 5). The criteria selection thus displays the filtered results in plots. The wearable data quality management feature operates entirely based on user-defined settings. For example, when a user sets the filter to include only days with a minimum wear time of 18 hours, VITAL displays only the data collected on those specific days (Figure 5). Operationalized metrics are also available for review (Figure S4 in Multimedia Appendix 3). In addition, the data export menu allows users to download the dataset with quality management applied. For this study, the entire process was documented with screenshots (Figures S5 and S6 of Multimedia Appendix 3).

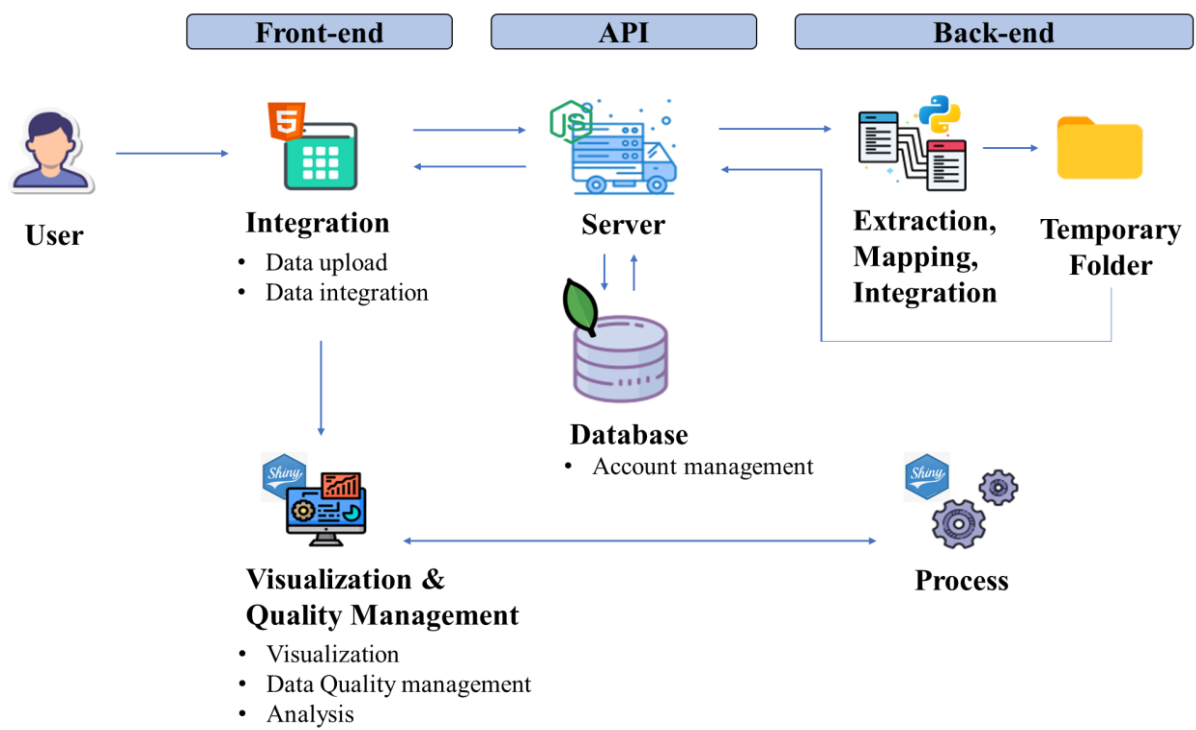

**Figure 4. The VITAL data processing pipeline.**

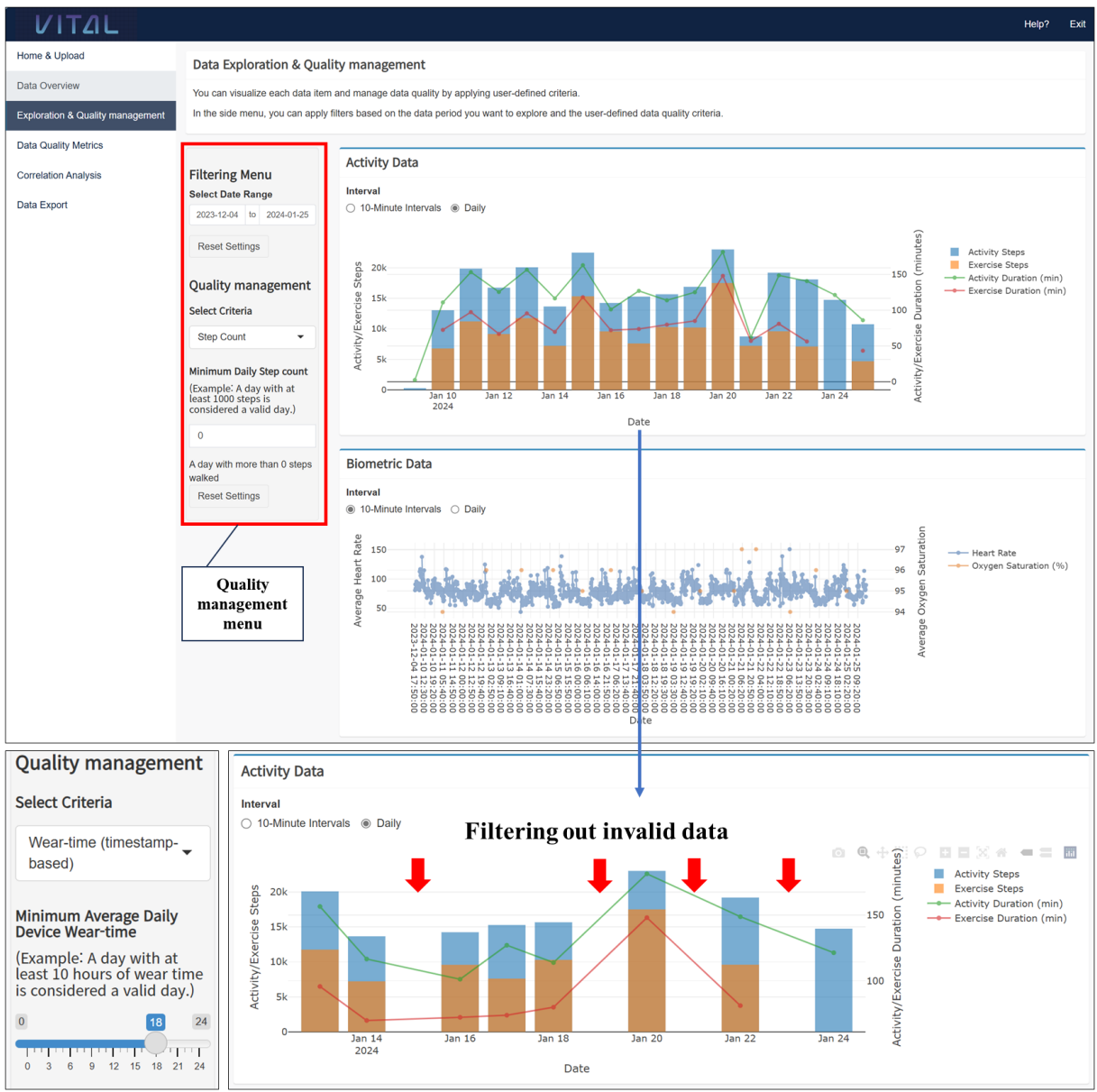

**Figure 5. Interactive data visualization and quality management screen.**

## Usability Evaluation

A total of seven clinicians (four MDs and three nurses) participated in the usability evaluation between February and March, 2024. Participant evaluations were carried out over three approximately one-hour sessions (Figure 3). The backgrounds of each participant are provided in Table 4.

**Table 4. Usability evaluation participant demographic characteristics.**

| Participants | 1 | 2 | 3 | 4 | 5 | 6 | 7 |
|---|---|---|---|---|---|---|---|
| Gender | Female | Male | Female | Female | Male | Male | Male |
| Age (years) | 28 | 35 | 32 | 42 | 38 | 33 | 43 |
| Occupation | Nurse | Nurse | Physician | Physician | Nurse | Physician | Physician |
| Education Level | Bachelor | Bachelor | Medical Doctor (MD) | Doctor of Medicine (PhD) | Bachelor | Graduate School of Medicine | Medical Doctor (MD) |
| Work Experience | 5 years in Hemato-Oncology, Tertiary Hospital | 6 years in Anesthesiology, Tertiary Hospital | 1 year in Internal Medicine, Tertiary Hospital | 4 years in Surgery, Tertiary Hospital; 18 years in other roles | 6 years in ICU, Tertiary Hospital | 2 years in Internal Medicine, Tertiary Hospital | 15 years in Internal Medicine, including other settings |
| Current Workplace | Hemato-Oncology Department | Anesthesiology Department | Student | Public Health Center | IT Department | Internal Medicine Department, Tertiary Hospital | Local Internal Medicine Clinic |

## Task completion

Table 5 details task completion results. The first task involved uploading and integrating data, and was successfully completed by all participants. The average time to task completion was approximately 2 minutes and 30 seconds. Only one error was observed across all participants, occurring in the case of one individual out of seven. For the second task, the average time for exploration was 4 minutes and 53 seconds. Two participants skipped over two specific menus (Participant 2: quality management and analysis; Participant 7: overview and analysis). All participants responded to the verification question regarding whether they performed the data quality management process correctly.

**Table 5. Results of task completion.**[a-e]

| # of task | Usability metrics | Participant 1 | Participant 2 | Participant 3 | Participant 4 | Participant 5 | Participant 6 | Participant 7 | Summary Metrics |
|---|---|---|---|---|---|---|---|---|---|
| 1 [a] | Task success | Success | Success | Success | Success | Success | Success | Success | 100% success |
| | Extra clicks [c] | 1 | 0 | 0 | 0 | 1 | 0 | 0 | - |
| | Error rate [d] | 0 | 0 | 0 | 0 | 1 | 0 | 0 | - |
| | Time per task (sec) | 127 | 94 | 148 | 295 | 279 | 60 | 45 | Mean 149.7sec (SD 100.4) |
| 2 [b] | Number of skipped menus | 0 | 2 | 0 | 0 | 0 | 0 | 2 | - |
| | Time per task (sec) | 447 | 120 | 259 | 391 | 296 | 435 | 105 | Mean 293.3sec (SD 141.4) |
| | Post-task accuracy [e] | Correct | Correct | Correct | Correct | Correct | Correct | Correct | 100% correct |

[a] 1: Data preprocessing (data upload and integration)
[b] 2: Data exploration and quality management. As task 2 was an open-ended exploration process, task success, extra clicks, and error rate were not measured.
[c] Extra clicks: The number of clicks exceeding the minimum required to complete the task (measured by menu or button clicks).
[d] Error rate: The number of times an incorrect item or button was clicked or selected.
[e] Post-task accuracy: The verification question to confirm whether the participant correctly performed the data quality management process.

After reviewing authentic wearable data, feedback was requested from the clinicians. In general, participants acknowledged that while the amount of exercise was satisfactory, there was a need to increase its intensity. However, they stated that relevant interventions would be needed to address the lack of sufficient sleep. The clinicians' feedback is summarized below in Textbox 1; Multimedia Appendix 4).

**Textbox 1. Clinician Feedback Based on Hypothetical Patient Data.**

> **Data Summary:** The sample data consisted of a total of 19 days, with an average daily step count of 15,466 and an average sleep duration of 4 hours.
> - "The patient walked a lot, but looking at the heart rate, it doesn't seem like a high-intensity workout. So, it looks like the exercise wasn't that intense." [Participant 1]
> - "The patient has been exercising well and should be able to maintain this effort going forward. But since their sleep quality isn't great, I'll ask about their sleep environment and give them tips on how to improve it." [Participant 4]
> - "I would recommend increasing the exercise routine. Since the patient is not completely unable to sleep, I will provide guidance on managing sleep hygiene effectively." [Participant 6]

### *Unified Theory of Acceptance and Use of Technology (UTAUT) Survey*

The score distribution of performance expectancy, effort expectancy, and participant intention to use was examined (Figure 6). The average score was 4.2 (SD 0.89), for performance expectancy, 3.96 (SD 1.03) for effort expectancy, and 4.14 (SD 0.92) for intention to use.

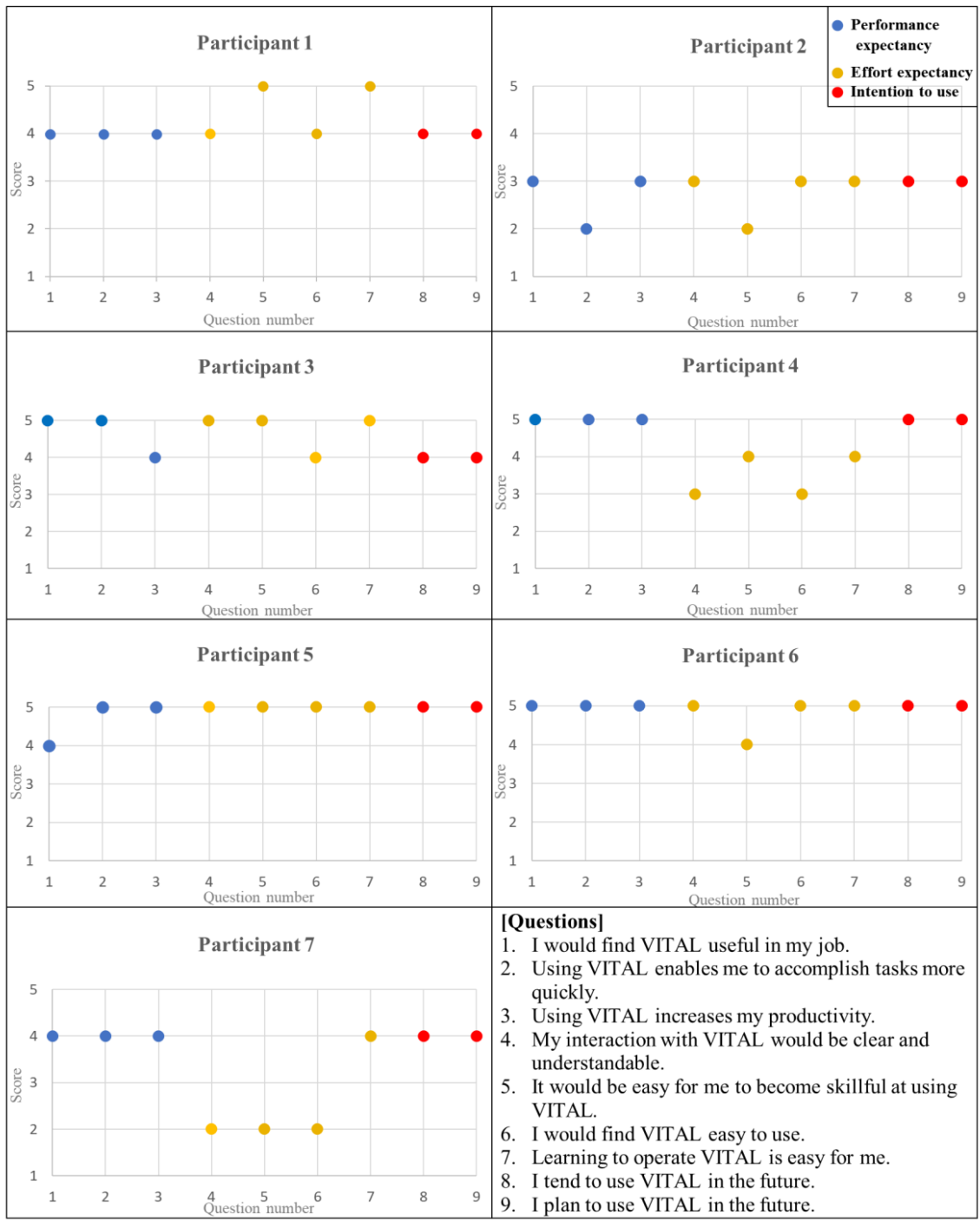

**Figure 6. Score distribution for performance expectancy, effort expectancy, and intention to use.**

## One-on-One Unstructured Interviews

Table 6 summarizes the collection of unstructured interview responses. Participants suggested simplifying the screen layout and menus. Specifically, participants noted that the location of the menu tabs at the top of the screen made them less noticeable (2 participants). There were no additional comments made regarding system functionality. However, there was a notable request for the inclusion of additional data items, with electrocardiogram (ECG) and dietary information being the most frequently requested (4 participants).

**Table 6. Participant feedback on VITAL prototype: usability, data quality, and clinical use.**

| Categories | Summary of Responses |
|---|---|
| **Interface (menu and screen layouts, etc.)** | **Strengths**:<br>• Fonts, overall color scheme, and layout were satisfactory<br>• Data exploration screen was intuitive and easy to understand, and satisfactory<br>• Data correlation analysis screen was viewed an interesting feature<br>**Areas for Improvement:**<br>• Screen layout should be simplified<br>• There were too many menus<br>• Placing the menu tabs at the top made them less noticeable<br>• Information related to data quality should be collapsible |
| **System Functionality** | All participants find current features sufficient.<br>• Download function for correlation analysis results |
| **Additional Information Needs** | • Electrocardiogram (ECG)<br>• Dietary information (intake amount/nutrition)<br>• Blood pressure<br>• Respiratory rate<br>• Others (Body Mass Index, stress, apnea status, snoring, and blood glucose, etc.) |
| **Data Filtering Criteria** | **Comment:**<br>• Step count varied with the patient's condition, making its use in clinical settings inappropriate as it was specific to the study context<br>• Outliers could be removed or specifically filtered for analysis.<br>• Having filtering criteria may be helpful; however, it is not |

| Categories | Summary of Responses |
|---|---|
| | considered an essential feature |
| **Clinical Applicability and Required Support** | All participants would use the VITAL prototype. **Necessary Support:** <br>• Staff support was deemed necessary for data collection and uploading to the system <br>• Integration with EMR was considered important; Simplification of tool menus and processes was identified as a need <br>• Device-related considerations (device provision, lightweight and simple to use) were highlighted <br>• Patient data collection process was viewed as a major barrier |
| **Recommendation to Other Healthcare Professionals** | Six out of seven participants would recommend the VITAL prototype. <br>• Helpful for patient care <br>• Biometric data would be useful in clinical practice |

All participants agreed on the necessity of data filtering criteria for quality management. While there was a general consensus on the adequacy of the filtering criteria provided by VITAL (wear time, step count), the step count criteria were considered inadequate due to their sensitivity to variations in patients' health status. Some participants also suggested additional criteria, such as the removal of outliers.

All participants expressed a willingness to use the VITAL prototype in the future and additionally noted that technical support for data collection and upload processes would be necessary (5 participants). With the exception of one participant, the majority voiced a willingness to recommend the VITAL prototype to their colleagues.

## DISCUSSION

### Principal Findings

Significant barriers continue to hinder the effective utilization of PGHD in clinical practice even though its potential is widely recognized. This study undertook comprehensive efforts to process wearable data with the aim of improving clinical workflows. A data processing pipeline was established, which integrated heterogeneous data sources into a standardized dataframe, and developed a system for the efficient exploration of large volumes of data. Furthermore, data quality management was implemented using data filtering methods and operational metrics. Usability evaluations provided by healthcare professionals offered valuable insights into VITAL's potential.

This study selected smartwatches and activity trackers from Samsung, Apple, Fitbit, and Xiaomi to examine and visualize the wearable data generated by these devices. Analysis of data generated by devices from these four manufacturers revealed substantial differences in file format, data structure, timestamps, and measurement units across devices. A data integration process was conducted to structurally and semantically standardize the data to facilitate efficient processing in the visualization module. Due to the variability in data collection intervals and recording durations among items, all data were integrated into a uniform 10-minute interval to reduce data distortion and enhance the efficiency of trend visualization. However, we acknowledge the decision to use a 10-minute interval was made based on limited experiments. To establish the most appropriate integration frequency, future research should rigorously validate these intervals. Moreover, device heterogeneity continues to present significant challenges to the standardization and interoperability of wearable data, which are both critical for application in a clinical setting. Addressing these issues requires collaborative efforts to ensure seamless data integration between manufacturers for consensus and standardization [6,21,23,42].

The VITAL prototype was evaluated by clinicians who with tertiary general hospital experience. The clinicians were willing to adopt VITAL for use in their clinical practices. Six of the seven participants indicated a willingness to recommend VITAL to their colleagues. One physician specifically expressed eagerness to use the tool immediately for patient consultations. Even without training, all participants found VITAL easy to use. Using VITAL, the clinicians were able to explore sample wearable data and provide useful feedback to the hypothetical patient based on the data. Although further research is necessary, these findings suggest that VITAL can help clinicians gain valuable insights into their patients' health and develop tailored treatment strategies for them [9].

Concerns about the quality of PGHD for clinical use remain despite the technological progress made in data accuracy [17,43,44]. Indeed, the importance of ensuring data quality was emphasized by all of the clinicians who participated in the VITAL evaluation. Nevertheless, there is currently no consensus on managing the quality of wearable health data for clinical use [23]. Currently, as seen in other studies, the VITAL prototype manages data quality using step count and wear time filters [33,34,38-40], However, step count data can vary significantly by factors such as patient age, health condition, and other social circumstances [45-47], which may limit the validity of step count as a quality criterion. Therefore, further research is needed to identify robust approaches to manage the quality of wearable health data for clinical practice.

## Limitations

Our study focused on devices that are widely used in South Korea. However, our study did not encompass all available devices or data types. Continuous efforts should be made to develop data processing methods applicable to more diverse device types and data items since each device generates data in different formats, time intervals, and measurement units. The integration interval was set to 10 minutes in this study, but further validation is needed to verify the optimal frequency of this interval. Although several data quality criteria from prior research studies were applied to VITAL, using step count as a key criterion proved difficult for broad applicability. Therefore, it is necessary to establish criteria that can be more widely applied in clinical practice. Future research should focus on refining a version of VITAL by addressing the aforementioned limitations and assessing its effectiveness through real-world clinical implementation.

## Comparison with Prior Work

We developed VITAL, an open-source prototype designed to process wearable health data for clinical applications. While previous efforts have created tools to process and visualize wearable health data for usability [48-52], VITAL distinguishes itself in several key ways. It integrates diverse data types from multiple manufacturers into a unified data table, facilitating comprehensive analysis. Additionally, VITAL provides interactive visualization features, enabling users to identify meaningful patterns efficiently.

A key innovation of VITAL is its emphasis on data quality management. It incorporates metrics for assessing completeness, recency, and plausibility of data, facilitating systematic evaluation of wearable data quality. It also includes wear-time-based filters, which enable users to derive reliable insights from high-quality data. While these features require further refinement, this study marks significant progress toward operationalizing data quality concepts for wearable health data. For example, a prior study introduced the completeness metric to evaluate data quality in a web-based tool for research application [53]. VITAL builds on this foundation by expanding the completeness metric to consider both the variety of data types and their temporal coverage. Moreover, it introduces additional dimensions of data quality, such as recency and plausibility, further enhancing the utility and reliability of wearable health data in clinical and research applications.

**Conclusions**

Wearable devices are increasingly capable of measuring a wide range of health data with improved accuracy. As a result, these devices play an important role in personal health management, and their use in the mHealth field is expanding. In clinical settings, wearable data can provide valuable insights for identifying patient health issues and making treatment decisions. In this context, the significance of the VITAL prototype is that it introduces the first comprehensive pipeline for the integration, visualization, and quality management of wearable data. The VITAL prototype provides a robust foundation for assessing data reliability through quality management as well as facilitates efficient data visualization. Positive feedback from study participants indicates that the VITAL prototype holds substantial potential for clinical application. Further implementation studies are needed to validate its effectiveness in real-world clinical settings.

2040. doi:10.1093/jamia/ocac166